\documentclass[aps,prb,showpacs,twocolumn,amsmath,amssymb,superscriptaddress]{revtex4}
\usepackage{graphicx}
\usepackage{Jonasmacros}
\usepackage{bm}
\usepackage{mathptmx}

\usepackage[usenames,dvipsnames]{color}

\begin{document}
\title{Inelastic electron tunneling spectroscopy at local defects in graphene}
\author{J. Fransson}
\email{Jonas.Fransson@physics.uu.se}
\affiliation{Department of Physics and Astronomy, Uppsala University, Box 516, SE-751 21 UPPSALA, Sweden}

\author{J. -H. She}
\affiliation{Theoretical Division, Los Alamos National Laboratory, Los Alamos, New Mexico 87545, USA}

\author{L. Pietronero}
\affiliation{Dipartimento di Fisica, La Sapienza Universitˆ di Roma, P.le A. Moro 5, 00185 Rome, Italy}
\affiliation{CNR-ISC, Via dei Taurini 19, 00185 Rome, Italy}

\author{A. V. Balatsky}
\affiliation{Theoretical Division, Los Alamos National Laboratory, Los Alamos, New Mexico 87545, USA}
\affiliation{Center Integrated Nanotechnology, Los Alamos National Laboratory, Los Alamos, New Mexico 87545, USA}
\affiliation{NORDITA, Roslagstullsbacken 23, SE-106 91\ \ STOCKHOLM, Sweden}

\begin{abstract}
We address local inelastic scattering from vibrational impurity adsorbed onto graphene and the evolution of the local density of electron states near the impurity from weak to strong coupling regime. For weak coupling the local electronic structure is distorted by inelastic scattering developing peaks/dips and steps. These features should be detectable in the inelastic electron tunneling spectroscopy, $d^2I/dV^2$, using local probing techniques. Inelastic Friedel oscillations distort the spectral density at energies close to the inelastic mode. In the strong coupling limit, a local negative $U$-center forms in the atoms surrounding the impurity site. For those atoms, the Dirac cone structure is fully destroyed, that is, the linear energy dispersion as well as the V-shaped local density of electron states is completely destroyed. We further consider the effects of the negative $U$ formation and its evolution from weak to strong coupling. The negative $U$-site effectively acts as local impurity such that sharp resonances appear in the local electronic structure. The main resonances are caused by elastic scattering off the impurity site, and the features are dressed by the presence of vibrationally activated side resonances. Going from weak to strong coupling, changes the local electronic structure from being Dirac cone like including midgap states, to a fully destroyed Dirac cone with only the impurity resonances remaining.
\end{abstract}
\date{\today}
\maketitle

\section{Introduction}
Graphene has been at the center of attention ever since its was first synthesized and studied for its unique physical properties.\cite{novoselov2004,geim2007,katsnelson2007,castroneto2009,vozmediano2010} While its properties are interesting on its own, increasing effort is also being directed towards modifications of  graphene. Functionalization of graphene has been achieved by depositing e.g. H atoms, thus, creating graphane\cite{elias2009} which is an insulator with a band-gap of the order of 3 | 6 eV. Chemical acid treatment may lead to vacancy formation in graphene,\cite{jafri2010} which tends to increase its conductivity due to a metallic-like density of electron states (DOS) in the vicinity of the vacancies.\cite{carva2010} The role of single and double vacancies in graphene has also been theoretically investigate, showing the emergence of midgap states.\cite{wehling2007} 

Modifications of electronic states and of the excitation spectrum of a given material is crucial for a more efficient functionalization. Examples of spectroscopies that are sensitive to electronic properties are, e.g. photoemission and photoabsorption techniques which give access to the bulk electronic structure, and local scanning techniques such as atomic force microscopy\cite{stowe1997,gross2009} and scanning tunneling microscopy\cite{binnig1982} (STM). They are employed for studies of spatial inhomogeneities\cite{lang2002} and local spectral properties.\cite{gomes2007}

By studying the response to defects in/on the material important spectroscopic information can be accessed.\cite{balatsky2006} For local probes this is a particularly fruitful strategy since it is relatively easy to move the probe on and off the defect. One thus can achieve comparable measurements of the perturbed and unperturbed material on one and the same sample. Through such an approach effects from potential, charge, and magnetic scattering can be measured from both elastic\cite{hasegawa1993,crommie1993} and inelastic\cite{grobis2005} point of view. Lately is has become routine to measure the inelastic electron tunneling spectrum (IETS) using STM.

In this paper we apply same logic to IETS in graphene. We calculate the local density of electron states (LDOS) for electrons in tight-binding honeycomb lattice which is used as a model for graphene. The main results are:
\begin{enumerate}
\item In the weak coupling limit and using perturbation theory, the LDOS near the local vibrational impurity exhibits a kink and logarithmic singularity at the vibrational mode $\omega_0$. The spectral density is significantly modified at energies near the vibrational mode. We predict those features to be observable in IETS experiment using local scanning techniques.

\item For strong coupling, the atoms surrounding the vibrational impurity forms negative $U$ centers such that the system can be considered as a single impurity problem, however, the impurity is effectively spatially extended. The LDOS is formed by a series of delta peaks forming a single band at negative energy. The result is universal in the sense that it is independent of the band structure of the conduction electrons, see also She et al. \onlinecite{she2013}.

\item By coupling the atoms influenced by the vibrational impurity to the surrounding lattice, we study the evolution of its LDOS from weak to strong coupling using a many-body approach. In the weak coupling regime, the Dirac cone is modified by the introduction of elastic resonances, surrounded by inelastic resonances, suggesting that the negative $U$-center effectively acts as local impurity. Here, the meaning of the local Dirac cone is related to the local energy dispersion and LDOS which deviate from being linear and V-shaped in a neighborhood of the impurity. In the single impurity case, we find that all, empty, singly, and doubly occupied, states are populated with a finite probability, which suggests the formation of local Cooper pair. For increasing coupling, the set of elastic and inelastic peaks move to lower energies below the Fermi level leaving a strongly asymmetric cone structure around the Fermi level. The Dirac cone is eventually fully destroyed in the strong coupling limit, leaving two resonances which are broadened by the inelastic resonances.
\end{enumerate}

The present work has some similarities and difference with previous study of inelastic signatures generates by local vibrational defect located on surface of topological insulator, Ref. \onlinecite{she2013}, and we point out a few differences which justifies the present study. The first apparent difference is that our present model for graphene is based on a discrete real space lattice instead of a continuum model, which implies that the exact location of the defect plays a role in the expected real space IETS imaging. This assumption also implies that the negative $U$-center may be induced at one or more sites simultaneously, depending on whether the vibrational defect couples to one or more C atoms in the graphene lattice. A second important difference is that we here have to deal with spinors of pseudo-spin, in which the entries depend on the sublattice instead of the electron spin. Thus, here we do not expect to obtain any possibility for magnetic contrast in the IETS. Finally, in our present study we treat the evolution from weak to strong coupling using a different approach by means of which we verify the main characteristics for each regime as compared to the case of topological insulators. Using this approach, however, we capture some central feature of the many-body (self-energy) aspects induced in the vicinity of the vibrational defect, and get direct access to electron number of the negative $U$-center. Moreover, due to the discreteness and bibpartite structure of the graphene lattice, the effective coupling between C atoms near the vibrational impurity cannot be removed by canonical transformation, see Sec. \ref{sec-strong}, which implies that the electronic and vibrational degrees of freedom cannot be separated without any (further) approximation.

The paper is organized as follows. First we set up the model for the graphene lattice and the vibrational impurity in Sec. \ref{sec-probe}. Then, we move on to discussing the weak coupling limit using a $T$-matrix approach in Sec. \ref{sec-weak} and the evolution from the weak to strong coupling limit using a many-body approach in Sec. \ref{sec-strong}. We finally conclude the paper in Sec. \ref{sec-conclusions}.

\section{Probing the inelastic scattering}
\label{sec-probe}
We describe the graphene sheet by the nearest neighbor interaction model
\begin{align}
\Hamil_0=&
	-t\sum_{\av{mn}\sigma}\Psi_{m\sigma}^\dagger\sigma_x\Psi_{n\sigma},
\end{align}
where the pseudo-spinor $\Psi_{m\sigma}=(\ac{m}\ \bc{m})^t$ contains the operators $a$ ($b$) which annihilate electrons in the $A$ ($B$) sub-lattice, and where $t$ is the hopping parameter.

By depositing molecular defect, e.g. CO, on the graphene sheet, a local vibrational mode can be introduced. Generically, the molecular vibrations cause non-static lattice distortions. Here, we specifically consider plaquette position of the vibrational impurity. An diatomic molecule may, for example, be located inside a hexagon in a straight up but slightly tilted position.\cite{biplab} The existence of six equivalent positions that the molecule can assume, due to the sixfold rotational symmetry of the hexagon, may cause molecular rotations, which generate local lattice distortions that can be described in term of a local bosonic mode coupling to the electronic density at the nearest C atoms. Stretching and breathing modes may also be envisioned, and especially if the molecule is off-centered within the hexagon. Thus, the coupling may be symmetric or asymmetric to the near carbon atoms. Here, we shall consider both possibilities since the latter can be reduced to effective single and double site interactions.

\begin{table}[b]
\caption{Vectors in momentum space connecting the lattice points.}
\label{tab1}
\begin{tabular}{l|ccc}
\hline\hline
 & 1 & 2 & 3
\\\hline
$\bfdelta_m$ & $a(\sqrt{3},1)/2$ & $-a(\sqrt{3},-1)/2$ & $-a(0,1)$
\\
$\bfdelta_{Am}$ & $a(0,1)$ & $-a(\sqrt{3},1)/2$ & $a(\sqrt{3},-1)/2$
\\
$\bfdelta_{Bm}$ & $-a(\sqrt{3},-1)/2$ & $-a(0,1)$ & $a(\sqrt{3},1)/2$
\\
\hline\hline
\end{tabular}
\end{table}

We, thus introduce $\omega_0B^\dagger B$, where $B^\dagger$ creates a vibron (local bosonic mode) at the energy $\omega_0$, for the local vibrational mode at the position $\bfR_0$. We describe its coupling to the nearest C atoms by
\begin{align}
\Hamil_\text{ep}=&
	\sum_{m\sigma}
		\Psi_{m\sigma}^\dagger
		\bflambda(\bfr_m)
		\Psi_{m\sigma}
		Q,\\
&\bflambda(\bfr_m)=
		\begin{pmatrix}
			\lambda_A(\bfr_m) & 0 \\
			0 & \lambda_B(\bfr_m)
		\end{pmatrix},
\nonumber
\end{align}
where $\lambda_{A/B}(\bfr_m)=\sum_{n=1}^3\lambda_{A/Bn}\delta(\bfR_0-\bfr_m+\bfdelta_{A/Bn})$, with $\bfdelta_{nA/B}$ are defined in Table \ref{tab1}, whereas $Q=B+B^\dagger$ is the vibrational displacement operator. Here, the coupling parameters $\lambda_n\neq\lambda_{n'}$ in general. While, in principle, the hopping parameter for the nearest neighbor interaction should be renormalized by the presence of the local vibrations, we neglect this effect here in order to keep the discussion as simple and transparent as possible.

Going over to momentum space via e.g. $\ac{m}=N^{-1/2}\sum_\bfk\ac{\bfk}e^{i\bfk\cdot\bfr_m}$, where $N$ denotes the number of C atoms in the $A$ sublattice, and analogously for the operators on the $B$-sublattice, we can write
\begin{align}
\Hamil_0=&
	\sum_{\bfk\sigma}\phi(\bfk)\adagger{\bfk}\bc{\bfk}+H.c.,
\end{align}
where the potential $\phi(\bfk)=-t\sum_{m=1}^3\exp{(i\bfk\cdot\bfdelta_m)}$ such that $\phi(\bfk+\bfK_\pm)\approx\pm v_Fk\exp\{\pm i(\pi/3-\varphi)\}$. Here, the vectors $\bfdelta_m$ are given in Table \ref{tab1}, $v_F=3at/2$, $\tan\varphi=k_y/k_x$ and $k=|\bfk|$, whereas $\bfK_\pm=\pm\bfK=\pm2\pi(\sqrt{3}/3,1)/3a$. The electron-vibron interaction Hamiltonian is in momentum space written as
\begin{align}
\Hamil_\text{ep}=&
	\sum_{\bfk\bfk'\sigma}
	\Psi_{\bfk\sigma}^\dagger
	\bflambda(\bfk,\bfk')
	\Psi_{\bfk'\sigma}
	Q,
\end{align}
where $\bflambda(\bfk,\bfk')=\diag{\lambda_A(\bfk,\bfk')\ \lambda_B(\bfk,\bfk')}{}$ and $\lambda_{A/B}(\bfk,\bfk')=\sum_m\lambda_{A/B}(\bfr_m)\exp{[-i(\bfk-\bfk')\cdot\bfr_m]}/N$.

\section{Weak coupling and $T$-matrix}
\label{sec-weak}
\begin{figure}[t]
\begin{center}
\includegraphics[width=0.99\columnwidth]{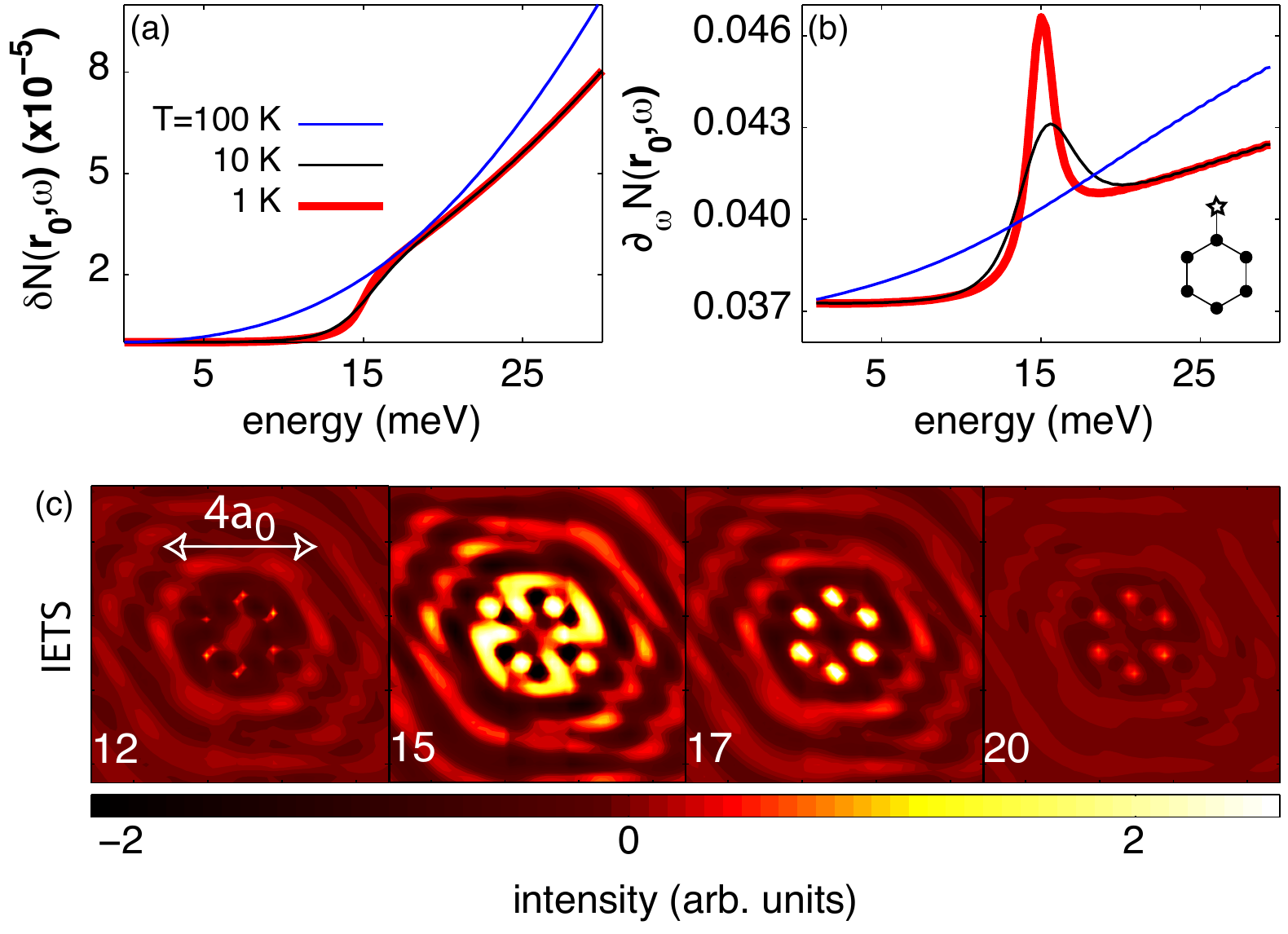}
\end{center}
\caption{(Color online) (a) Change in the local DOS, corresponding to $dI/dV$, and (b) its energy dependent derivative (IETS), corresponding to $d^2I/dV^2$,  at $\bfr_0=\bfR_0+2\bfdelta_{1A}$ (star in the inset) in the weak coupling limit, for different temperatures $T=100,\ 10,\ 1$ K and vibrational mode $\omega_0=15$ meV, for uniform coupling to the nearest C hexagon. (c) Sequence of IETS maps as function of energy, from left to right $\omega=12,\ 15,\ 17,\ 20$ meV, using $T=10$ K and spatial broadening $\Gamma=2a_0/5$. We have added an intrinsic broadening of $0.8$ meV in the potential $\bfV_{mn}$.}
\label{fig1}
\end{figure}

We study the effect of the a weak vibrational impurity by perturbation theory, which is valid for $\lambda_{A/Bn}/t\ll1$. The dressed graphene Green function (GF) $\bfG(\bfk,\bfk';z)=\av{\inner{\Psi_\bfk}{\Psi_{\bfk'}^\dagger}}(z)$, suppressing the spin indices, can be calculated in terms of the Dyson equation
\begin{align}
\bfG(\bfk,\bfk')=&
	\delta(\bfk-\bfk')\bfG_0(\bfk)
	+\bfG_0(\bfk)\sum_{\bfkappa}\bfSigma(\bfk,\bfkappa)\bfG(\bfkappa,\bfk'),
\end{align}
where
\begin{align}
\bfG_0(\bfk;z)=&
	\frac{1}{z^2-|\phi(\bfk)|^2}
	\begin{pmatrix}
		z & \phi(\bfk) \\
		\phi^*(\bfk) & z
	\end{pmatrix}
\end{align}
is the bare graphene GF, whereas the self-energy is given by
\begin{align}
\bfSigma(\bfk,\bfk';z)=&
	\sum_{mn}\int
		e^{-i\bfk\cdot\bfr_m}
			\bflambda(\bfr_m)
			\bfV_{mn}(z)
			\bflambda(\bfr_n)
		e^{i\bfk'\cdot\bfr_n}.
\end{align}
Here, the potential $\bfV_{mn}(z)=i\beta^{-1}\sum_\nu D(z_\nu-z)\bfG(\bfr_m,\bfr_n;z_\nu)$, where we sum over Bosonic frequencies $z_\nu=i2\nu\pi/\beta$, $\nu\in\mathbb{Z}$, $\beta=1/k_BT$, and where we have introduced the local Boson GF $D(z)=\av{\inner{Q}{Q}}(z)$. In the weak coupling limit, we replace both dressed GFs in $\Sigma$ by their bare correspondences, using $D_0(z)=2\omega_0/(z^2-\omega_0^2)$. Accordingly, the GF is cast in $T$-matrix form in real space
\begin{subequations}
\begin{align}
\bfG(\bfr,\bfr')=&
	\bfG_0(\bfr-\bfr')
\nonumber\\&
	+\sum_{mn}\bfG_0(\bfr-\bfr_m)\bfT(\bfr_m,\bfr_n)\bfG_0(\bfr_n-\bfr'),
\\
\bfT(\bfr_m,\bfr_n)=&
	\Bigl(\delta(\bfr_m-\bfr_i)-\bfG_0(\bfr_m-\bfr_i)\bfV_{ij}\Bigr)^{-1}\bfV_{jm},
\end{align}
\end{subequations}
with the bare real space GF given by
\begin{align}
\bfG_0(\bfR)=&
	\frac{2\pi\omega}{iD_c^2}
	\biggl(
		H_0^{(1)}\bigg(\frac{\omega R}{v_F}\biggr)
		\sigma_0\cos\bfK\cdot\bfR
		-iH_1^{(1)}\bigg(\frac{\omega R}{v_F}\biggr)
\nonumber\\&\times
			\Bigl(
				\sigma_x\sin\theta_R\sin\bfK\cdot\bfR
				+i\sigma_y\cos\theta_R\cos\bfK\cdot\bfR
			\Bigr)
		\biggr).
\end{align}
Here, $H_n^{(1)}(\omega)$ is the $n$th Hankel function of the first kind, whereas $\sigma_i$, $i=x,y,z$, are Pauli matrices and $\sigma_0$ is the identity matrix. Here, also $\bfR=\bfr-\bfr'$, $\theta_R=\phi_R+\pi/6$, $\tan\phi_R=(r_y-r_y')/(r_x-r_x')$, whereas $D_c^2=4\pi\rho v_F^2$, with surface density $\rho=S/N=k_c^2/4\pi$ ($S$ graphene area; $k_c=2\sqrt{2\sqrt{3}\pi}/3a$ large momentum cut off).\cite{peres2006} We comment here that the Fourier transform $\bfG_0(\bfR)=\int\bfG_0(\bfk)d\bfk/(2\pi)^2$ is convergent and does not depend on any specific details of the large momentum cut off $k_c$, something which has been discussed in Ref. \onlinecite{kogan1211} for the case of Ruderman-Kittel-Kasuya-Yosida (RKKY) interaction in graphene and pertains to our discuss as well. The cut off $k_c$ is introduced in order to maintain a physical finite density $\rho$.

Integration around $\pm\bfK$, yields the retarded potential (with obvious notation and $x_{mn}=p|\bfR_{mn}|/v_F$)
\begin{widetext}
\begin{align}
\bfV^r_{mn}(\omega)=&
	\frac{2}{D_c^2}
	\bflambda(\bfr_m)
	\sum_{s=\pm1}
	\int_0^{D_c}
		\biggl(
			\frac{1+n_0-f(p)}{\omega-sp-\omega_0+i\delta}
			+\frac{n_0+f(p)}{\omega-sp+\omega_0+i\delta}
		\biggr)
\nonumber\\&\times
		\biggl(
			J_0(x_{mn})\sigma_0\cos\bfK\cdot\bfR_{mn}
			-isJ_n(x_{mn})
			[
				\sigma_x\sin\theta_{mn}\sin\bfK\cdot\bfR
				+i\sigma_y\cos\theta_{mn}\cos\bfK\cdot\bfR
			]
		\biggr)
	pdp
	\bflambda(\bfr_n).
\label{eq-Vdef}
\end{align}
\end{widetext}
Here, $f(x)$ is the Fermi distribution function whereas $n_0=n(\omega_0)$ is the Bose distribution function at $\omega_0$.

We remark here that adatoms may be a source for scattering processes with large momentum transfer which would cause an intervalley coupling. For instance, in momentum space the electron-vibron Hamiltonian has the from $\sum_{\bfk\bfk'}\Psi^\dagger_\bfk\bflambda(\bfr)\exp[-i(\bfk-\bfk')\cdot\bfr]\Psi_{\bfk'}$, which we can write as $\sum_{\bfk\bfp}\Psi^\dagger_{\bfk+\bfp}\bflambda(\bfr)\exp[-i\bfp\cdot\bfr]\Psi_\bfk$. The latter form explicitly indicates intervalley coupling (large $\bfk+\bfp$). However, as we employ the $T$-matrix expansion, we do not have to worry about intervalley coupling since we use the former expression for the electron-vibron Hamiltonian, in which the momentum summations are separated, hence, the valleys are decoupled. This, thus, justifies that we integrate around $\pm\bfK$ only.

\begin{figure}[b]
\begin{center}
\includegraphics[width=0.99\columnwidth]{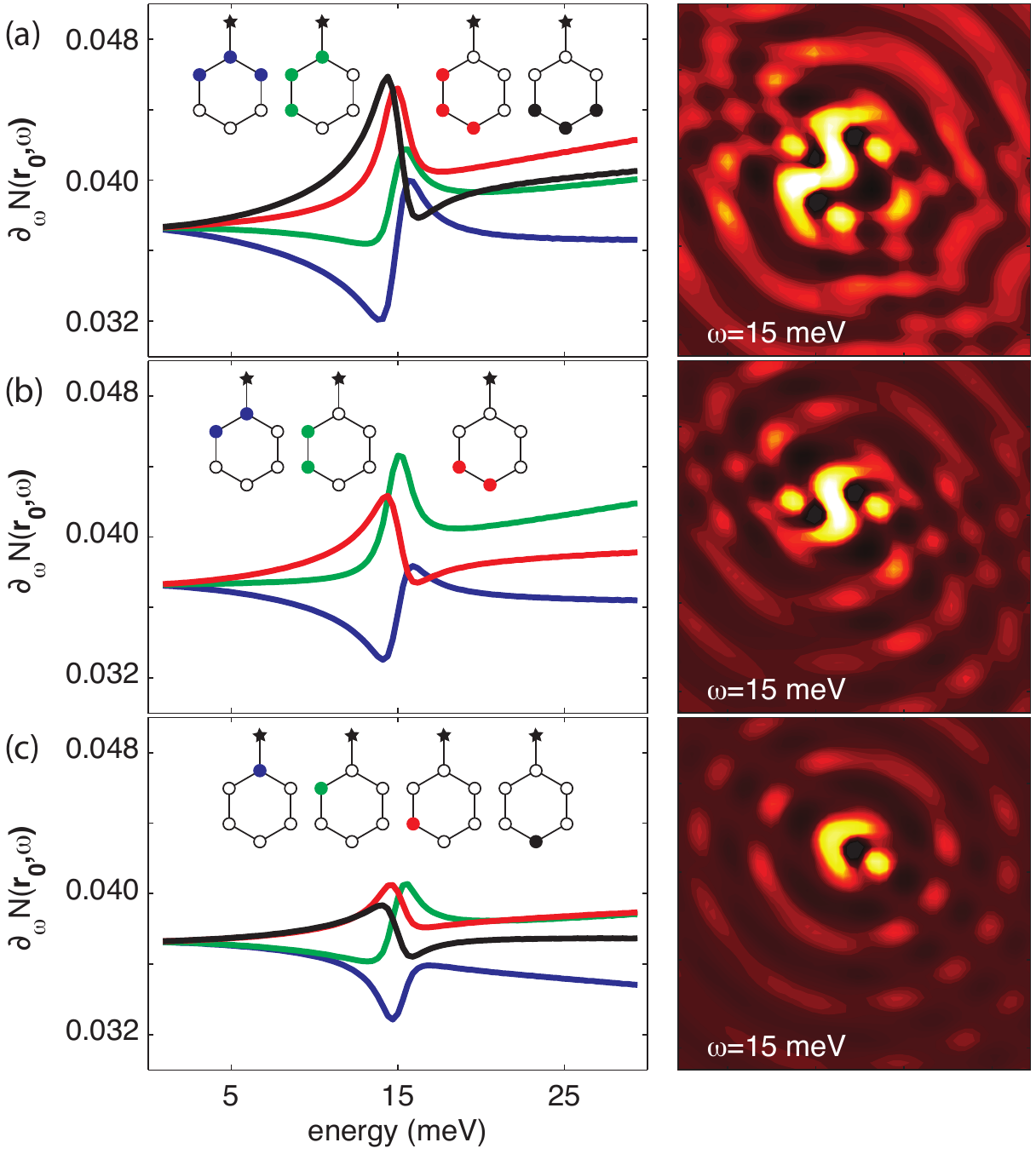}
\end{center}
\caption{(Color online) Change in the local IETS at $\bfr_0$ in the weak coupling limit, for different asymmetric configurations with coupling to (a) three, (b) two, and (c) one C atom in the nearest neighbor hexagon, and distance from point of measurement, as indicated in the upper insets. Left panels show the corresponding IETS maps at $\omega=\omega_0$. Parameters as in Fig. \ref{fig1} (c).}
\label{fig2}
\end{figure}

The electronic structure around the vibrational impurity is modified at energies near the inelastic mode $\pm\omega_0$, where a kink and peak/dip is created due to the inelastic scattering off the vibrational center. Using uniform coupling to the hexagon surrounding the vibrational impurity, in Fig. \ref{fig1} we plot the correction to the local density of electron states (LDOS), panel (a), and its energy derivative (IETS), corresponding to $d^2I/dV^2$, panel (b), at $\bfR_\tip-\bfR_0=a(0,2)$, for different temperatures. The LDOS shows non-trivial structure at the vibrational mode which are more apparent in the IETS as peaks around $\omega_0=15$ meV. Similar features are also predicted for the case of IETS signatures in d-wave superconductors\cite{balatsky2003} and in topological insulators,\cite{she2013} as well as for simple metals both for vibrational\cite{fransson2007} and magnetic imputiry.\cite{fransson2012}

The corresponding real space mapping of the IETS is displayed in Fig. \ref{fig1} (c) for energies below, near, and above $\omega_0$. For energies below and above $\omega_0$, the presence of the local vibrations generate low contrast, while the contrast grows substantially larger for energies around $\omega_0$. We expect that the presence of the vibrations generates sufficiently large variations in the IETS, i.e. $d^2I/dV^2$, to be visible in an experimental set-up.

We complete the weak coupling picture by also plotting the IETS signatures for asymmetric coupling in Fig. \ref{fig2}, assuming (a) three, (b) two, and (c) one, C atom being coupled to the vibrational impurity. As one may expect, the IETS signal is stronger when more C atoms are coupled to the vibrational impurity. We also plot different distances between the measuring point at $\bfr_0=\bfR_0+2\bfdelta_{1A}$ and the atom(s) that are coupled to the vibrational impurity, clearly showing the oscillatory behavior that is expected due the inelastic Friedel oscillations (see insets of the figure, and Fig. \ref{fig1}).

\section{Evolution from weak to strong coupling regime}
\label{sec-strong}
We here depart from the $T$-matrix approximation and consider the evolution of features from weak to strong coupling, i.e. for coupling parameter $\lambda_{A/B}/t\gtrsim1$, using many-body theory. First, we decouple the Fermionic and Bosonic degrees of freedom near the vibrational impurity using the small polaron transformation,\cite{lang1962} that is, constructing the Hamiltonian $\widetilde\Hamil=e^S\Hamil e^{-S}$ with
\begin{align}
S=&
	i\frac{P}{\omega_0}
	\sum_{m\sigma}
		\Psi_{m\sigma}^\dagger
		\bflambda(\bfr_m)
		\Psi_{m\sigma}
		,\
		P=(-i)(B-B^\dagger).
\end{align}
We can write the resulting model according to
\begin{align}
\widetilde\Hamil=&
	-t\sum_{\av{mn}\sigma}
		\Psi_{m\sigma}^\dagger
		e^{-i\bflambda(\bfr_m)P/\omega_0}
		\sigma_x
		e^{i\bflambda(\bfr_n)P/\omega_0}
		\Psi_{n\sigma}
\nonumber\\&
	+\omega_0B^\dagger B
	-\biggl(
		\sum_{m\sigma}
			\Psi_{m\sigma}^\dagger
				\tilde\bflambda(\bfr_m)
			\Psi_{m\sigma}
	\biggr)^2
		,
\label{eq-tildeH}
\end{align}
where $\tilde\bflambda(\bfr_m)=\bflambda(\bfr_m)/\sqrt{\omega_0}$.

The above expressions are valid for all couplings $\lambda_{A/B}(\bfr_m)$, and clearly shows that the presence of the inelastic scattering center gives rise to an attractive interaction for the electrons residing on the atoms surrounding the vibrational center. The appearance of the electron-vibron couplings in the first term of $\widetilde\Hamil$ is due to the fact that $S$ does not commute with $\adagger{i}\bc{j}$.

\subsection{Strong coupling limit}
\label{ssec-strong}
Before we discuss the evolution of the electronic structure from the weak to strong coupling regime, we first consider a few observations about the strongly coupled system. In the strong coupling limit, the system reduces to a single impurity problem, with the difference to the conventional impurity problem in that here the impurity is constituted of up to six C atoms around the vibrational defect, depending on the symmetry/asymmetry of the coupling. For asymmetric coupling such that the vibrational impurity effectively couples to one C atom, the system reduces to a single site problem in which the Fermionic ground states can be written, for example, $\ket{2}=a^\dagger_{1\up}a^\dagger_{1\down}\ket{}$, where $\ket{}$ denotes the empty state, assuming that the vibrational impurity couples to atom $n=1$ in the $A$ sublattice without loss of generality. The excited states are $\ket{\sigma}=a^\dagger_{1\sigma}\ket{}$ and $\ket{0}=\ket{}$, and the Fermionic energy spectrum can be written $E_\nu=-(\nu\tilde\lambda_{A1})^2$, $\nu=0,1,2$. The energy gain for doubly occupied site is evident from this result hence we expect this \emph{local} attraction to play a major role in inducing pairing correlations in graphene due to local bosonic mode.

With the above observations in mind, we write the Hamiltonian of the negative $U$-site as
\begin{align}
\widetilde\Hamil_\text{imp}=&
	-\tilde\lambda^2_{A1}
	\sum_\sigma
		(
			1+a^\dagger_{1\bar\sigma}a_{1\bar\sigma}
		)
		a^\dagger_{1\sigma}a_{1\sigma}.
\label{eq-Himpa1}
\end{align}
In terms of the eigenspectrum of the negative $U$-center, we write $\ac{1}=\X{0\sigma}{}+\sigma\X{\bar\sigma2}{}$, where $\X{pq}{}\equiv\ket{p}\bra{q}$ denotes the transition from state $\ket{q}$ to $\ket{p}$ and the factor $\sigma\equiv\sigma^z_{\sigma\sigma}$. We can, thus, write
\begin{align}
\widetilde\Hamil_\text{imp}=&
	E_0\X{00}{}+E_1\sum_\sigma\X{\sigma\sigma}{}+E_2\X{22}{}.
\end{align}

\begin{figure}[t]
\begin{center}
\includegraphics[width=0.99\columnwidth]{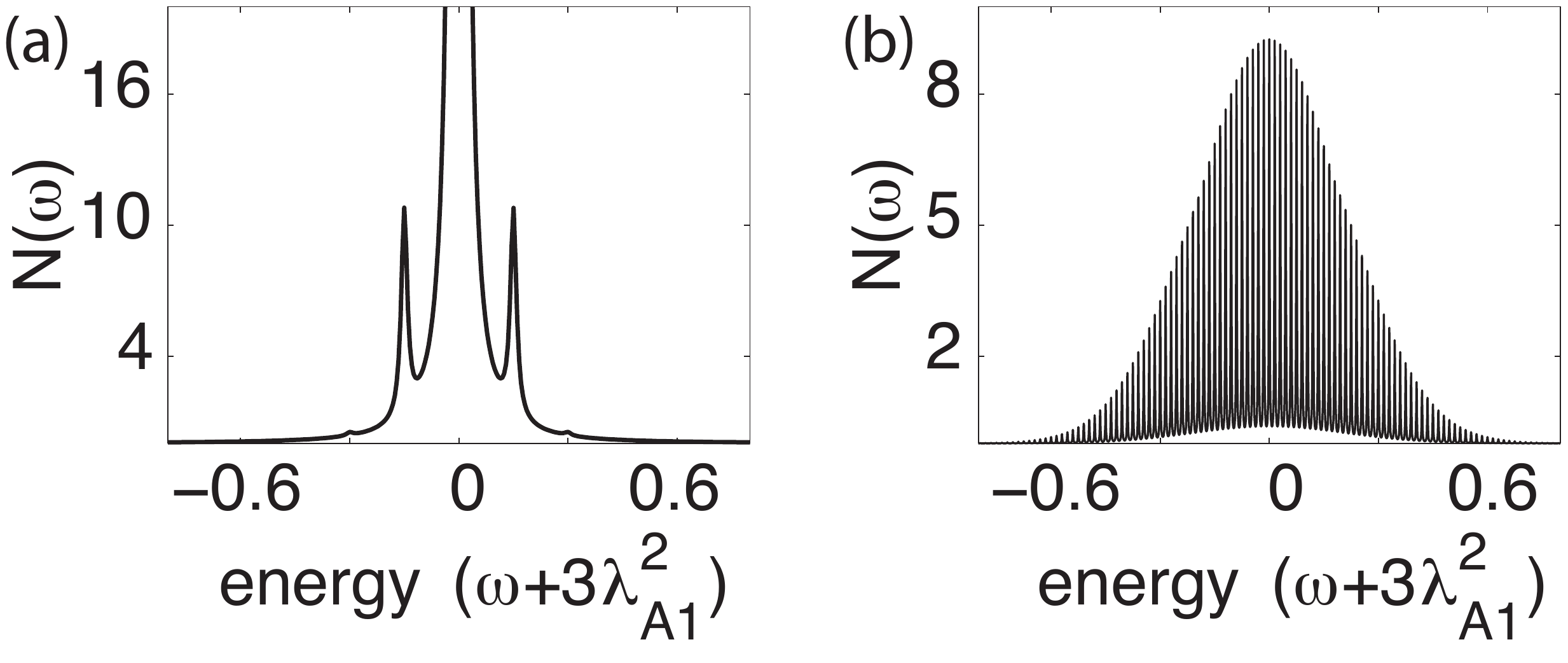}
\end{center}
\caption{DOS for the single site problem in the strong coupling/atomic limit using $\omega_0=15$ meV and $\lambda/D_c=2\cdot10^{-2}$ for (a) $T=10$ K and (b) $T=100$ K.}% Inset in panel (a) shows a close up of the plot in the main panel.}
\label{fig3}
\end{figure}

The spectrum of the single site is determined through the GF $\tilde{\bfG}(t,t')=\bfG_{\sigma\sigma'}(t,t')\bfF(t,t')$, where $\bfG(t,t')=\eqgr{\Psi_{n\sigma}(t)}{\Psi^\dagger_{n\sigma'}(t')}$ is the electronic GF and $\bfF_n(t,t')=\{F_{n\alpha\beta}(t,t')\}_{\alpha,\beta=A,B}$,
\begin{align}
F_{n\alpha\beta}(t,t')=&
	\av{\calX_{n\alpha}(t)\calX^\dagger_{n\beta}(t')}_\text{vib},
\end{align}
is the average over the bosonic degrees of freedom. Here,
\begin{align}
\calX_{n\alpha}(t)=&
	e^{i\omega_0B^\dagger Bt}e^{i\lambda_\alpha(\bfr_n) P/\omega_0}e^{-i\omega_0B^\dagger Bt},\
	\alpha=A,B.
\end{align}
Following the procedure lined out in e.g. Ref. \onlinecite{mahan}, we calculate the generalized function ($\tau=t-t'$)
\begin{align}
F_{n\alpha\beta}(t,t')=&
	\exp\Biggl\{
		-\frac{1}{2\omega_0^2}
		\Biggl[
			(1+2n_0)
			\biggl(
				\lambda_\alpha(\bfr_n)+\lambda_\beta(\bfr_n)
			\biggr)^2
\nonumber\\&
			-2\lambda_\alpha(\bfr_n)\lambda_\beta(\bfr_n)
			\biggl(
				(1+n_0)(1+e^{-i\omega\tau})
\nonumber\\&
				+n_0(1+e^{i\omega\tau})
			\biggr)
		\Biggr]
	\Biggr\},
\end{align}
giving the Fourier transformed GF
\begin{align}
\tilde\bfG_{\sigma\sigma'}^r(\omega)=&
	e^{-(1+2n_0)
		[\lambda_\alpha^2(\bfr_n)+\lambda_\beta^2(\bfr_n)]/2\omega_0^2}
\nonumber\\&\times
	\sum_n
		I_n(\tilde\omega_0)
		e^{n\beta\omega_0/2}
		\bfG_{\sigma\sigma'}^r(\omega-n\omega_0),
\end{align}
where $\tilde\omega_0=2\lambda_\alpha(\bfr_n)\lambda_\beta(\bfr_n)\sqrt{n_0(1+n_0)}/\omega_0^2$, and where $I_n(x)$ is the modified Bessel function. Thus, for the single site problem given by Eq. (\ref{eq-Himpa1}), the electronic ground state is in the atomic limit given by the GF
\begin{align}
G_{\sigma\sigma'}^r(\omega)=&
	\delta_{\sigma\sigma'}
	\biggl(
		\frac{1-\av{\adagger{1}\ac{1}}}{\omega+\tilde\lambda_{Aa}^2+i\delta}
		+\frac{\av{\adagger{1}\ac{1}}}{\omega+3\tilde\lambda_{Aa}^2+i\delta}
	\biggr),
\end{align}
$\delta>0$. Setting $\av{\adagger{1}\ac{1}}=1$, which corresponds to the double occupied configuration, we reproduce the analogous spectrum found in Ref. \onlinecite{she2013} for vibrational impurity on surface of topological insulator, i.e. a series of sharp peaks centered around the two-Fermion energy $-3\tilde\lambda_{A1}^2$. This is shown in Fig. \ref{fig3} for (a) $T=10$ K and (b) $T=100$ K, also showing that more inelastic side peaks become activated with increasing temperature, as expected. Similar conclusions hold for all our considered cases with $N=1,\ldots,6$ C atoms coupling to the vibrational impurity, with Fermionic ground state consisting of $2N$ electrons.

\subsection{Evolution from weak to strong coupling}
\label{ssec-evolution}
Considering further the single site problem, now in presence of the surrounding lattice, we write the transformed lattice Hamiltonian as
\begin{align}
\widetilde\Hamil_0=&
	\Hamil_0+\widetilde\Hamil_T,
\end{align}
where the coupling between the negative $U$-center and the lattice is given by
\begin{align}
\widetilde\Hamil_T=&
	\sum_{\bfk\sigma}
		t_\bfk
			\Bigl(
				1-e^{-i\lambda_{A1}P/\omega_0}
			\Bigr)
			(\X{\sigma0}{}+\sigma\X{2\bar\sigma}{})
			\bc{\bfk}
		+H.c.,
\end{align}
with $t_\bfk=-t\sum_{n=1}^3e^{i\bfk\cdot(\bfr_1+\delta_n)}/\sqrt{N}$, such that $t_{\bfk\pm\bfK}\approx\pm v_Fke^{\pm i(\pi/3-\varphi)+i\bfk\cdot\bfr_1}/\sqrt{N}$. The negative $U$-center, hence, couples to the surrounding lattice with an effective hybridization $\tilde t_\bfk$ which is renormalized by the momentum $P$ of the local bosonic mode.

We capture the evolution from the weak to strong coupling limit by solving the equation of motion for the many-body operator GF $\mathbb{G}_{a\bar{b}}(z)=\av{\inner{\X{a}{}}{\X{\bar{b}}{}}}(z)$, for the transitions $a,b=(0\sigma),(\sigma2)$, self-consistently in mean-field approximation under the self-consistency condition that the occupation numbers $N_0+\sum_\sigma N_\sigma+N_2=1$. The occupation numbers are calculated using\cite{fransson2005}
\begin{subequations}
\label{eq-N}
\begin{align}
N_0=&
	-\frac{1}{\pi}\im\sum_\sigma\int[1-f(\omega)]\mathbb{G}^r_{0\sigma\sigma0}(\omega)d\omega,
\\
N_\sigma=&
	-\frac{1}{\pi}
	\im
	\int[
		f(\omega)\mathbb{G}^r_{0\sigma\sigma0}(\omega)
		+[1-f(\omega)]\mathbb{G}^r_{\sigma22\sigma}(\omega)
	]d\omega,
\\
N_2=&
	-\frac{1}{\pi}\im\sum_\sigma\int\mathbb{G}^r_{\sigma22\sigma}(\omega)d\omega.
\end{align}
\end{subequations}
Due to the inherent spin-degeneracy and absence of a coupling between the spin-channels, the GF reduces to a $2\times2$-matrix equation. To second order in $t_\bfk$ and $\tilde\lambda$, the result is given in terms of the retarded GF
\begin{align}
\mathbb{G}^r(\omega)=&
	\biggl(
		\omega-\bfDelta
		-\mathbb{P}\Sigma(\omega)(1+\sigma_x)
	\biggr)^{-1}
	\mathbb{P},
\label{eq-mfG}
\end{align}
where $\bfDelta=\diag{\Delta_1 \ \Delta_2}{}$, $\Delta_n=E_n-E_{n-1}$, $\mathbb{P}=\diag{P_1\ P_2}{}$, $P_1=N_0+N_1/2$, $P_2=N_1/2+N_2$, $N_1=\sum_\sigma N_\sigma$, whereas the self-energy $\Sigma(\omega)=\sum_{n=1,2}\Sigma^{(n)}(\omega)$ is given by
\begin{subequations}
\label{eq-S}
\begin{align}
\bfSigma^{(1)}(\omega)=&
	-2\omega
	\biggl[
		1+\biggl(\frac{\omega}{D_c}\biggr)^2
		\biggl(
			2\log\frac{D_c}{|\omega|}
			+i\pi\sign\omega
		\biggr)
	\biggr]
	,
\label{eq-S1}
\\
\bfSigma^{(2)}(\omega)=&
	-4\pi\frac{f(\omega)}{D_c^2}\frac{\omega}{\omega^2-\omega_0^2}
		\frac{\omega+\tilde\lambda^2}{\omega+\tilde\lambda^2/2}
		\omega^3\sign\omega
	.
\label{eq-S2}
\end{align}
\end{subequations}
The contribution $\Sigma^{(1)}$ account for fluctuations on and off the negative $U$-center, essentially caused by the presence of the surrounding lattice, showing a cubic correction to the LDOS. The second contribution, $\Sigma^{(2)}$, is generated by fluctuations on and off the negative $U$-center due to the coupling between the local vibrational mode and the Fermionic degrees of freedom.

Equation (\ref{eq-mfG}) using the self-energies in Eq. (\ref{eq-S}) should be solved self-consistently, however, we can make a few observations on the expected behavior of the electronic structure. For weak coupling, the bare excitations $E_\nu=-(\nu\tilde\lambda_{A1})^2\rightarrow0$. Thus, for low energies, such that $\omega/D_c\ll1$, we can neglect the self-energy $\Sigma^{(2)}$ and approximate the first self-energy by $\Sigma^{(1)}\approx-2\omega$. Then, the denominator of $\mathbb{G}^r$ is given by
\begin{align}
3\omega^2-2\omega\sum_{n=1}^2\Delta_n+\prod_{n=1}^2\Delta_n
=3(\omega-\Delta_+)(\omega-\Delta_-),
\label{eq-simplepoles}
\end{align}
where $\Delta_\pm=-(4\mp\sqrt{7})\tilde\lambda^2/3$. This is found by observing that $\Delta_1=-\tilde\lambda^2$ and $\Delta_2=-3\tilde\lambda^2$, such that $\sum_n\Delta_n=-4\tilde\lambda^2$ and $\prod_n\Delta_n=3\tilde\lambda^4$. Here, we have, moreover, used that $P_1+P_2=N_0+\sum_\sigma N_\sigma+N_2=1$, by charge conservation, along with $P_n\approx1/2$, c.f. Fig. \ref{fig5}.

As the coupling is increased, the non-linear components in the self-energies play an increasingly important role for the positions of the poles, such that we cannot any longer make use of Eq. (\ref{eq-simplepoles}).

\begin{figure}[t]
\begin{center}
\includegraphics[width=0.99\columnwidth]{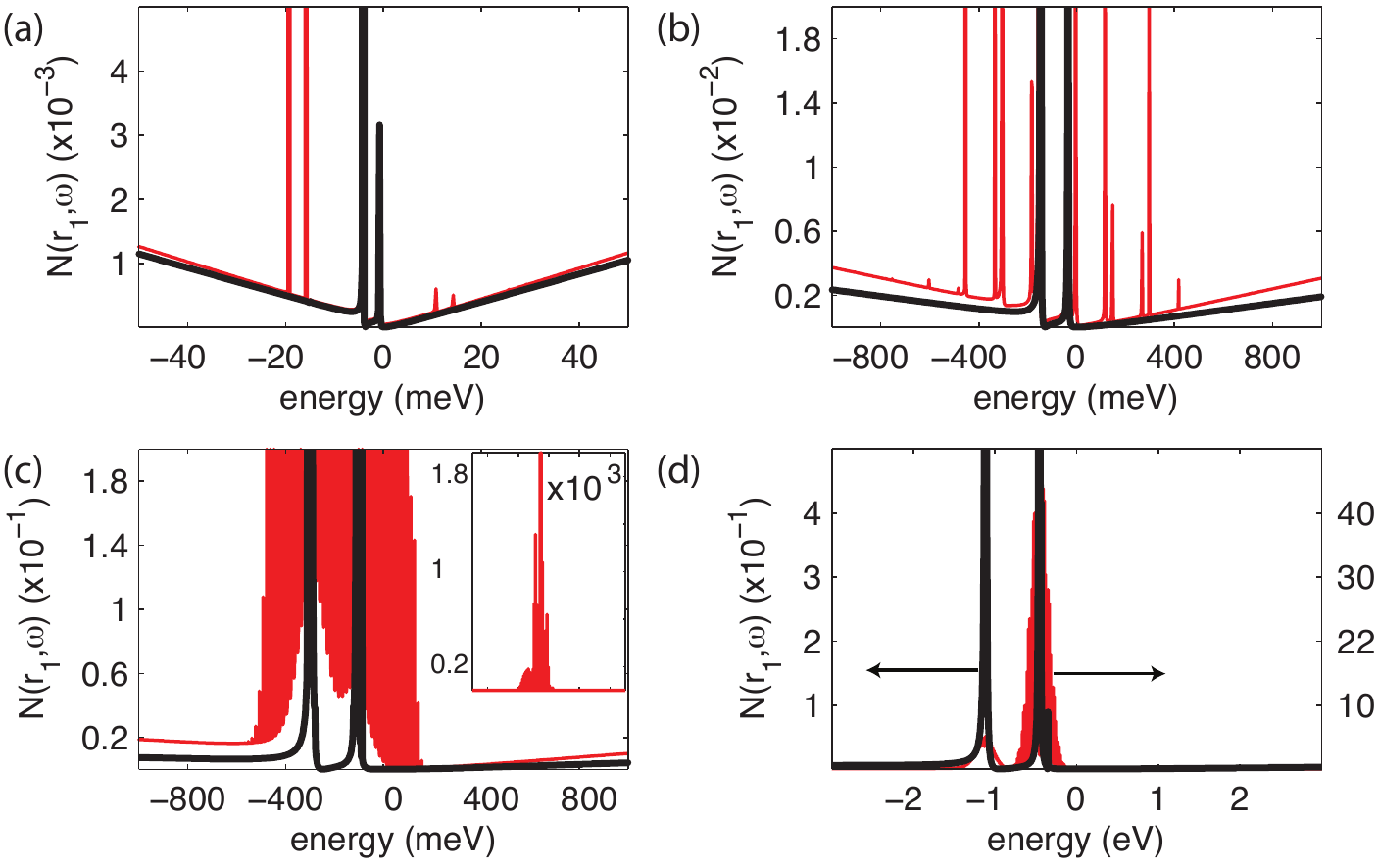}
\end{center}
\caption{(Color online) Evolution of the LDOS at the negative $U$-center from weak to strong coupling regime. Here, $\lambda/D_c=\{5\cdot10^{-4},\ 1\cdot10^{-3},\ 5\cdot10^{-3},\ 1\cdot10^{-2}\}$, $\omega_0=15$ meV, and $T=10$ K (bold/black) and $T=100$ K (faint/red). The inset in panel (c) shows the full LDOS at $T=100$ K.}
\label{fig4}
\end{figure}

In Fig. \ref{fig4} (a) | (d) we plot the evolution of the LDOS on the negative $U$-center from weak to strong coupling regime for low (bold/black) and high (faint/red) temperatures. The LDOS, $\rho(\omega)=-\tr\im\mathbb{G}^r(\omega)/\pi$, is obtained from solving Eqs. (\ref{eq-N}) and (\ref{eq-mfG}) self-consistently under the condition $N_0+\sum_\sigma N_\sigma+N_2=1$. In the weakly coupled system, panels (a), there are two main (elastic) peaks near the Fermi level, corresponding to $\Delta_\pm$, c.f. Eq. (\ref{eq-simplepoles}). For low temperatures there is tiny signature of a vibrational side peak at about $\omega=-\omega_0=-15$ meV. For higher temperatures, these vibrational signatures become more apparent, as one should expect since those modes are thermally activated.

For increasing coupling the main elastic features remain, however, shifted to lower energies. They become increasingly broadened since the level width is cubic function of the energy, c.f. Eq. (\ref{eq-S1}). Moreover, the presence of the vibrational side peaks also become more visible in the LDOS, even for low temperatures. In both  cases illustrated by panels (a) and (b), the coupling is weak enough to preserve the overall Dirac cone, apart from the presence of the resonances.

For even stronger coupling, panels (c) and (d), the Dirac cone is fully destroyed and only the peak features, caused by the elastic and inelastic scattering, remain. Finally, in the strong coupling limit, panel (d), there only appears a double peak structure, where the peaks correspond to the singly and doubly occupied states. For high temperatures, the vibrational side peaks effectively act as a thermal broadening of the main peaks. The discrepancy with the situation illustrated in Fig. \ref{fig3} can be understood from the fact that we here take into account fluctuations to both the singly and doubly occupied states, hence, there is a finite likelihood that even the singly occupied state becomes populated. This is typical feature of any many-body description, and it emphasizes the fact that the charge is partially distributed among the available states.

\begin{figure}[t]
\begin{center}
\includegraphics[width=0.99\columnwidth]{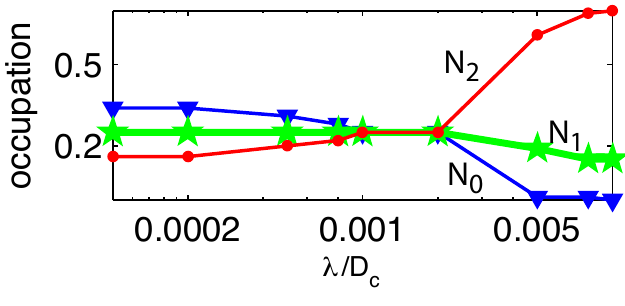}
\end{center}
\caption{(Color online) Evolution of the occupation numbers $N_0$ (triangles), $N_1$ (pentagrams), and $N_2$ (bullets), at the negative $U$-center from weak to strong coupling regime. Here, $\omega_0=15$ meV and $T=10$ K.}
\label{fig5}
\end{figure}

We finally comment on the evolution of the Fermionic state of the negative $U$-center from weak to strong coupling regime, represented in terms of the populations numbers $N_n$, $n=0,1,2$, c.f. Fig. \ref{fig5}. In the weakly coupled system, the energy of the single electron fluctuations $\Delta_n=-(n\tilde\lambda)^2+(n-1)^2\tilde\lambda^2=-(2n-1)\tilde\lambda^2$ lies below but close to the Fermi level, c.f. Fig. \ref{fig4} (a), (b), such that the system is open for fluctuations between the (four) states. This property is verified by the occupation numbers, in that all $N_n$, $n=0,1,2$ are finite. This suggests occurrence of local Cooper pair formation near the vibrational impurity, which will be the topic of a future publication.

In the strongly coupled limit, on the other hand, the set of elastic and inelastic transition energies are far below the Fermi level, c.f. Fig. \ref{fig4} (d), such that the the population number $N_0$ approaches zero. In other words, the negative $U$-center acquires a Fermionic ground state which is a mixture of the singly and doubly occupied states. The coupling between the negative $U$-center and the surrounding lattice, thus, generates a more intricate electronic structure than what is suggested by the atomic limit physics where the negative $U$-center is decoupled from the lattice.

In the intermediate regime, there is a cross-over regime, or possibly a phase transition, c.f. crossing of population numbers near $\lambda/D_c\gtrsim10^{-3}$ in Fig. \ref{fig5}, where the occupation numbers of the empty and doubly occupied states evolve monotonically decreasing and increasing, respectively, with the coupling strength $\lambda$, whereas the single Fermion state(s) remain constant.

It is, finally, worth mentioning that the attractive force indicated by Eq. (\ref{eq-Himpa1}) always have to be compared to the repulsive Coulomb forces present in the material. For the case of graphene, there is a controversy whether there is a significant contribution to the electronic structure caused by the Coulomb interaction, which is closely related to the question whether the ground state of graphene is in non-magnetic semi-metallic state or a anti-ferromagnetic insulating state.\cite{drut2009} While the latter seems to be favorable for suspended graphene, the former situation pertains to graphene deposited on a substrate which complies with our initial assumption. For this case, graphene is very well described by non-interacting electrons with negligible Coulomb interaction.

\section{Conclusions}
\label{sec-conclusions}
We have theoretically studied the effects of vibrational impurity adsorbed onto graphene, specifically the inelastic scattering properties. We find in the weak coupling regime, that the perturbed LDOS in the vicinity of the vibrational impurity acquires peaks/dips and steps at the energy of the vibrational mode. The spectral density distortions around the vibrational mode is spatially extended showing inelastic Friedel oscillations, in analogy with the findings for surfaces of metallic materials\cite{fransson2007,fransson2012,gawronski2011} and topological insulator.\cite{she2013}

By employing a many-body approach, we study the evolution from weak to strong coupling regime. In the weak coupling regime, an elastic mid-gap resonance emerge, surrounded by inelastic side resonances, at half the energy of the single electron fluctuations between the negative $U$-center and the surrounding lattice. Finite occupation of all Fermionic states, empty, singly, and doubly occupied state, on the negative $U$-site, near the vibrational impurity in the weakly coupled system, suggests local Cooper pair formation. The aspects of this physics will be the topic of a future publication.

For intermediate coupling strength the peak structure is severely distorted and pushed below the Fermi level, leaving a strongly asymmetric Dirac cone around the Fermi level. The Dirac cone is eventually destroyed in the strongly coupled regime, in which the electronic structure acquires a band formed by the collection of elastic and inelastic resonances.

We believe that our findings should be within the scope of present experimental local probing abilities using e.g. STM or atomic force microscopy.

\acknowledgements
JF acknowledges B. Sanyal for communicating unpublished results and J. -X. Zhu for fruitful discussions. The authors thank the Swedish Research Council, EU, and Nordita for support. JF further acknowledges the Wenner-Gren Foundation for travel support. Work at LANL was carried out under the auspices of the U.S. DOE under Contract No. DE-AC52-06NA25396 through the Office of Basic Energy Sciences, Division of Materials Science and Engineering, and the UC Research Fee Program.

\end{document}